\documentclass{article}
\usepackage[breakable]{tcolorbox}
\usepackage{parskip} %

\usepackage[T1]{fontenc}
\usepackage{mathpazo}

\usepackage{tabularx} %
\usepackage{amsmath}  %
\usepackage{bm} %
\usepackage[margin=1in,letterpaper]{geometry} %
\usepackage[final]{hyperref} %

\DeclareSymbolFont{cmletters}{OT1}{cmr}{m}{n}

\usepackage[dvipsnames]{xcolor}

\hypersetup{
	colorlinks=true,       %
	linkcolor=Blue,        %
	citecolor=Blue,        %
	filecolor=magenta,     %
	urlcolor=Blue         
}

\usepackage[
    style=numeric-comp,
    sorting=none,
    url=false,
    isbn=false,
    firstinits=true
]{biblatex}
\renewbibmacro{in:}{}
\AtEveryBibitem{
\clearlist{language}
\clearfield{month}
\ifentrytype{article}{
        \clearfield{note} 
        \clearfield{url}
        \clearfield{eprint}
    }{}
\ifentrytype{eprint}{
        \clearfield{doi} 
        \renewbibmacro*{eprint}{}
        \renewbibmacro*{addendum+pubstate}{}%
        \renewbibmacro*{note}{}%
        \renewbibmacro*{date}{}%
    }{}
}

\renewbibmacro*{finentry}{%
  \setunit{\finentrypunct\addspace}%
  \ifentrytype{eprint}{
    \printtext{\href{\thefield{url}}{\thefield{eprint}}}. (\printdate)
    }{}
  \finentry
}

\usepackage{multirow} %

\usepackage{esint}
\usepackage{amssymb}
\usepackage{gensymb}
\usepackage{float}
\usepackage{physics}
\usepackage{textgreek}

\usepackage{titling}

\title{Quantum (non)equivalence of dual massive\\$p$-form gauge theories}
\author{Christian Canete and Elden Loomes
\vspace{0.2cm} \\
\normalsize \itshape
Sydney Consortium for Particle Physics and Cosmology, \\
\normalsize  \itshape
School of Physics, The University of Sydney, NSW 2006, Australia \\
\normalsize \itshape \href{mailto:christian.canete@sydney.edu.au}{christian.canete@sydney.edu.au},\,\, 
\normalsize \itshape \href{mailto:elden.loomes@sydney.edu.au}{elden.loomes@sydney.edu.au}, }
\date{}

\begin{document}

\maketitle

\begin{abstract}
    Gauge theories of massive $p$-forms are connected by various dualities, which hold classically but may be broken at the quantum level. One example is the $BF$ theory of topologically coupled $p$- and $(d-p-1)$-forms in $d$ dimensions, where the coupling between forms results in a manifestly gauge invariant mass term for either form when the other is integrated out classically. We perform the path integral quantisation of this theory; by integrating out one of the forms, the resulting determinants are sensitive to the topology of spacetime, and counterterms must be introduced to renormalise their divergences. We compute these determinants in terms of the topological numbers of spacetime, showing explicitly how the quantum duality of the massive theories is broken on topologically non-trivial backgrounds. This is directly related to the quantum breaking of the massless duality between the form that was integrated out and the longitudinal modes of its partner. In particular, the difference of counterterms is proportional to the Euler characteristic of spacetime. The existence of gravitational instantons suggests that these dualities may be broken even in Minkowski space in the presence of topological fluctuations.
\end{abstract}

\tableofcontents

\section{Introduction}
Higher-form gauge theories are pervasive in theoretical physics, attractive as they are for their rich phenomenological implications in cosmology and particle physics. 
Massless 2-form gauge fields (known as Kalb--Ramond fields) in four dimensions were first explored in \cite{ogievetskii_notoph_1967,kalb_classical_1974} as a generalisation of forces between 1-dimensional strings. Furthermore, one can show that this 2-form can be reformulated into a theory of a pseudoscalar field \cite{gliozzi_supersymmetry_1977,chamseddine_n_1981,freund_kaluza-klein_1982,gasperini_pre-big-bang_1993,copeland_string_1995}.
Massive 2-forms non-minimally coupled to gravity have applications modelling different dynamics in inflation, with a typical signature involving an asymmetry in the primordial gravitational wave spectrum due to parity violation \cite{ford_inflation_1989,golovnev_vector_2008,kobayashi_gravitational_2009,capanelli_cosmological_2024,manton_kalb-ramond_2024,horii_ghost-free_2025,hell_kalb-ramond_2026}.

Massless 3-form gauge theories in four dimensions are intriguing in that they possess no propagating degrees of freedom, providing only an electrostatic-like field determined by the boundary conditions of the theory. In fact, they generate a source of negative pressure, and have been studied as possible candidate theories neutralising the cosmological constant \cite{hawking_cosmological_1984,duff_cosmological_1989,duncan_four-forms_1990}. In addition, massless 3-forms in a finite volume can exhibit an analogous Casimir effect in hadronic vacuums \cite{aurilia_quantum_2004}. Massive 3-form gauge theories, on the other hand, are classically dual to a massive pseudoscalar field i.e.~an axion, prompting interest as a possible UV completion. This has applications as a QCD axion candidate \cite{peccei_cp_1977} for the strong $CP$ problem \cite{dvali2005,dvali2022,sakhelashvili_consistency_2022}, and as a dark matter candidate originating from the vacuum energy of the 3-form \cite{ansoldi_vacuum_2001,klinkhamer_dark_2017,canete_emergent_2026}.

Higher form gauge fields may also have origins in supergravity theories.  After dimensional reduction of the $D=10$ or $11$ supergravity multiplet down to four dimensions, 2-forms and 3-forms naturally arise among the effective degrees of freedom \cite{cremmer_supergravity_1978,freund_dynamics_1980,freund_kaluza-klein_1982,fradkin_effective_1985,lukas_string_1997}. In a more familiar setting, it is well known that in ordinary gauge theories one can construct an Abelian Chern--Simons 3-form out of the gauge fields. Physically, the 3-form encodes the topological properties of the gauge field, such as the topological term proportional to $\theta_{\mathrm{QCD}}$ in quantum chromodynamics (QCD). Furthermore, it has been show that QCD in a non-trivial topology has implications for confinement in two dimensions and gluon bound states in four dimensions \cite{luscher_secret_1978,aurilia_u1_1980,gabadadze_modeling_1998}. In more recent studies, the Chern-Simons 3-form in the electroweak sector may in principle generate a mass gap through the breaking of the anomalous $U(1)_{B+L}$ symmetry, analogous to the $\eta'$ meson in QCD after chiral symmetry breaking, leading to a new degree of freedom in the Standard Model \cite{dvali_electroweak_2025,dvali__2025}.

It is well known that, at the classical level, there exists a duality between a massless $p$-form and a massless $(d-p-2)$-form, where $d$ is the number of spacetime dimensions. Similarly, there is a duality between a massive $p$-form and a massive $(d-p-1)$-form \cite{hjelmeland_duality_1997}.
Studies have been done to determine whether the duality of $p$-form gauge theories breaks at the quantum level. Indeed, it has been found that the duality between a massless 2-form gauge field and a scalar field is broken due to the one-loop gravitational trace anomaly; their difference is equal to the Gauss--Bonnet invariant, which may be non-zero for non-trivial topologies \cite{duff_quantum_1980}. Similarly, it has been shown that a massless 3-form gauge field also has a gravitational trace anomaly proportional to the Gauss--Bonnet invariant. There are, however, studies that challenge this claim, stating that the topological differences are trivial through the regularisation of the IR divergences of the zero modes \cite{siegel_quantum_1981,grisaru_energy-momentum_1984,buchbinder_quantization_1988}. In addition, it has been shown that non-minimal coupling to gravity may also break the $p$-form duality even at the classical level \cite{hell_duality_2022,felice_non-minimal_2025}.

An interesting question to ask is if this duality (breaking) extends to the massive gauge theories as well. Massive gauge theories have to be treated carefully, as adding a mass to a gauge field by hand explicitly breaks the gauge symmetry, so it must be introduced through a manifestly gauge-invariant manner e.g.~the Higgs mechanism \cite{anderson_plasmons_1963,englert_broken_1964,higgs_broken_1964} or the $BF$ formalism \cite{birmingham_topological_1991,leblanc_induced_1994,emery_two-dimensional_1998}. It has been shown that in trivial topologies where there are no zero modes, the duality in massive higher-form gauge theories holds exactly at the quantum level \cite{buchbinder_quantum_2008,kuzenko_effective_2021}. However, while \cite{kuzenko_effective_2021} shows that the any difference is at most topological, they do not compute this difference explicitly, nor show the role of zero modes in this breaking---our work aims to address this problem directly and show that the duality is not preserved in general when quantum effects are taken into account. %

In particular, we perform the path integral quantisation of the $BF$ theory, yielding a quantum effective action for a massive form. When one of the forms is dualised to provide longitudinal modes for the other, the mismatch of zero modes results in divergences, the counterterms for which probe the topology of spacetime. We show that this leads to different counterterms for the classically dual theories in the thermal partition functions on Schwarzschild and de Sitter spaces, breaking the duality. When applied to the Eguchi--Hansen spacetime \cite{eguchi_asymptotically_1978,eguchi_self-dual_1979}, thought to describe instantons in quantum gravity, our results suggest that the duality between massive theories may be broken even in the gravitational vacuum. 

The paper is structured as follows. In Section \ref{sec:dual_classic_0mass} we briefly summarise the classical duality of massless $p$-form gauge theories. In Section \ref{sec:dual_classic_massive} we describe the theory of massive $p$-forms which is manifestly gauge invariant through the $BF$ formalism. Using this framework, we then show the classical duality properties of massive $p$-forms. In Section \ref{sec:quantise} we quantise the $BF$ theory and show that the classically dual massive theories differ in the path integral by a topological factor. In Section \ref{sec:examples} we consider possible examples of this non-equivalence in a variety of cases. Finally, we summarise our findings in Section \ref{sec:conc}. %

\subsection{Notation and conventions}
\label{subsec:notations}
Let us first review the primary notations and conventions used throughout this paper. Let $\omega$ and $\chi$ be $p$-forms over a Riemannian manifold. %
We work in Euclidean signature, in $d$ dimensions. We denote the Hodge dual by $\star$ and the exterior derivative by $\mathrm{d}$. There is a inner product on $p$-forms, denoted
\begin{align}
    \langle\omega,\chi\rangle\equiv\int \omega\wedge\star \chi%
\end{align}
which is symmetric and duality invariant, $\langle\star \omega,\star\chi\rangle=\langle \omega,\chi\rangle$. Integration runs over the whole manifold unless otherwise noted. Define the codifferential $\delta$ as the adjoint of $\mathrm{d}$ such that $\langle\mathrm{d}\omega,\chi \rangle=\langle\omega,\delta\chi\rangle$ (up to boundary terms); acting on $p$-forms, we find $\delta =(-1)^{(d-k-1)(k+1)+k}\star \mathrm{d}\,\star$. Denote by $\Box\equiv-(\delta\mathrm{d}+\mathrm{d}\delta)$ the Hodge Laplacian which is negative semi-definite.\footnote{This choice of sign is opposite to the mathematical literature, but matches the usual Laplacian on scalar functions.} See Appendix \ref{app:note} for conventions regarding components of $p$-forms and normalisations.

\section{Duality of massless $p$-forms}
\label{sec:dual_classic_0mass}
First, we show the duality between a massless $p$-form and a $(d-p-2)$-form for $0\leq p \leq d-2$. We isolate the special case where $p=d-1$ in the end.

Suppose we have a massless $p$-form $A$, with gauge redundancy 
$A \rightarrow A+\mathrm{d}\omega$ where $\omega$ is a $(p-1)$-form. The gauge invariant Euclidean action for the massless $p$-form is just the kinetic term
\begin{align}
    S_p[A] = \frac{1}{2}\int F\wedge\star F = \frac{1}{2}\langle F,F\rangle ~, \label{eq:S[A]-forms}
\end{align}
where $F\equiv\mathrm{d} A$ is the field strength tensor of the $p$-form. Varying the action with respect to $A$ gives
\begin{align}
    \delta_A S_p[A] = \int \mathrm{d}\delta_A A\wedge\star F = \int\mathrm{d}\left(\delta_A A \wedge\star F\right) + (-1)^{p+1}\int \delta_A A\wedge\mathrm{d}\star F~.
\end{align}
Demanding that the field variation of the action is zero, we obtain the equation of motion $\mathrm{d} \star F = 0$, along with the boundary condition that variations of the $p$-form vanish when projected on the boundary, %
$\delta_A A|_{\mathrm{bdry}} = 0$ (other prescriptions are possible, e.g.~see \cite{duncan_four-forms_1990}). The Bianchi identity for the $p$-form, $\mathrm{d} F=0$, is automatically satisfied since $F$ is locally exact. 

Alternatively, we could begin by imposing the Bianchi identity, in which case the local exactness of the field strength defines $A$ up to gauge transformations automatically. The failure of this exactness is measured by the degree $(p+1)$ de Rham cohomology group $H^{p+1}(M;\mathbb{Z})$, over the spacetime $M$, which classifies the field strength into discrete classes differing by closed but not exact forms. Only in the trivial class does there exist a global smooth choice of $A$. The most famous example of this non-exactness is that of the Wu--Yang monopole \cite{wu_concept_1975,wu_dirac_1976}, where $H^2(M;\mathbb{Z})\cong \mathbb{Z}$ labels the discrete magnetic charges supported on a monopole defect. Hodge theory tells us that there is exactly one harmonic form, $\omega$ with $\Box\omega=0$, representing each cohomology class \cite{bott_differential_1982}.

Now, the equation of motion tells us that the Hodge dual of $F$ is also closed, and we can write $\star F = G$ where $G$ is a closed $(d-p-1)$-form. Like $F$, $G$ is then locally the derivative of some $(d-p-2)$-form $C$, up to gauge transformations $C \rightarrow C+\mathrm{d}\theta$. Thus, we can write $\star F$ as
\begin{align}
    \star F = G = \mathrm{d} C~. \label{eq:starF_massless}
\end{align}
The equation of motion and the Bianchi identity of the $p$-form in terms of the $(d-p-2)$-form are now $\mathrm{d} G = 0$ and $\mathrm{d} \star G = 0$, respectively. One can see that the duality transformation interchanges the equation of motion and Bianchi identity of the two massless theories. 

Substituting Eq.~\eqref{eq:starF_massless} into the action in Eq.~\eqref{eq:S[A]-forms} and using the fact that the inner product is Hodge dual-invariant in the Euclidean signature, we obtain the action for a $(d-p-2)$-form
\begin{align}
    S_{d-p-2}[C] = \frac{1}{2}\int G\wedge\star G = \frac{1}{2}\langle G,G\rangle~.
\end{align}
Thus, a massless $p$-form and a $(d-p-2)$-form are classically equivalent due to the duality of the theories on-shell.

As sanity check, we can also check the energy-momentum tensors of the theories and show that the two are equivalent. Indeed, the Euclidean energy-momentum tensor for the massless $p$-form is (briefly going back to component form)
\begin{align}
    T_{\rho\sigma}[A] = \frac{1}{p!}F_{\rho\mu_1\cdots \mu_p}{F_{\sigma}}^{\mu_1\cdots \mu_p} - \frac{1}{2(p+1)!}g_{\rho\sigma}F_{\mu_0\mu_1\cdots \mu_p}F^{\mu_0\mu_1\cdots \mu_p}~.
\end{align}
In the dual theory, where $\star F = G$, the equation can be written in component form as
\begin{align}
    F^{\mu_0\mu_1\cdots \mu_p} = \frac{1}{(r+1)!}\epsilon^{\mu_0\mu_1\cdots \mu_p\alpha_0\alpha_1\cdots \alpha_r}G_{\alpha_0\alpha_1\cdots \alpha_r}~, \quad G_{\alpha_0\alpha_1\cdots \alpha_r} = (r+1)\partial_{[\alpha_0}C_{\alpha_1\cdots \alpha_r]}~,
\end{align}
where $r \equiv d-p-2$, and $[\,\cdots]$ denotes normalised antisymmetrisation. By direct substitution it is possible to show that
\begin{align}
    T_{\rho\sigma}[A] = \frac{1}{r!}G_{\rho\alpha_1\cdots \alpha_r}{G_{\sigma}}^{\alpha_1\cdots \alpha_r} - \frac{1}{2(r+1)!}g_{\rho\sigma}G_{\alpha_0\alpha_1\cdots \alpha_r}G^{\alpha_0\alpha_1\cdots \alpha_r} = T_{\rho\sigma}[C]~,
\end{align}
proving their equivalency.

As another sanity check, we verify the duality of the massless $p$-forms by counting the number of propagating degrees of freedom. First, note that the equations of motion for a $p$-form in component form is
\begin{align}
    \partial_{\mu_0}\left(\sqrt{-g}\,\partial^{[\mu_0}A^{\mu_1\cdots \mu_p]}\right) = 0~.
\end{align}
Pick a time direction, indexed by zero and use Latin letters to denote spatial indices. We deduce that the time components of the gauge field $A_{0i_2\cdots i_{p}}$ are not dynamical, as their time derivatives do not enter in the equations of motion; rather, $A_{0i_2\cdots i_p}$ generate the Gauss constraint. Thus, the relevant dynamical components are the space-like components $A_{i_1\cdots i_p}$. They have gauge redundancy $A_{i_1\cdots i_p} \rightarrow A_{i_1\cdots i_p}+ \partial_{[i_1}\omega_{i_2\cdots i_p]}^{(p-1)}$ for $\omega_{i_2\cdots i_p}^{(p-1)}$ a $(p-1)$-form. Not every component of $\omega^{(p-1)}$ should be counted as removing a degree of freedom, however, because the transformation depends only on the exterior derivative of $\omega^{(p-1)}$, which then has its own gauge redundancy $\omega^{(p-1)}_{i_2\cdots i_p}\rightarrow \omega^{(p-2)}_{i_2\cdots i_p}+\partial_{[i_2}\omega^{(p-2)}_{i_3\cdots i_p]}$ for $\omega_{i_3\cdots i_p}^{(p-2)}$ a $(p-2)$-form. This has in turn its own redundancy, descending down to the scalar gauge transformation $\omega^{(1)}_{i_p}\rightarrow \omega^{(1)}_{i_p}+\partial_{i_p}\omega^{(0)}$ which has no further gauge redundancy. Therefore the number of propagating degrees of freedom after taking into account the full gauge redundancy is\footnote{The above relation can be proved by induction, which can conveniently be seen by noting that the sum $\sigma_p$ has a recursion relation
\begin{align}
    \sigma_p = \left(\begin{array}{c}
        d-1 \\
        p 
    \end{array}\right) - \sigma_{p-1}~. \nonumber
\end{align}}
\begin{align}
    \text{Propagating d.o.f.~(massless)} &= \left(\begin{array}{c}
        d-1 \\
        p
    \end{array}\right)  - \left\{\left(\begin{array}{c}
        d-1 \\
        p-1 
    \end{array}\right) - \left[\left(\begin{array}{c}
        d-1 \\
        p-2 
    \end{array}\right)-\cdots\right]\right\} \nonumber\\
    &= \sum_{n=0}^p(-1)^n\left(\begin{array}{c}
        d-1 \\
        p-n 
    \end{array}\right) = \left(\begin{array}{c}
        d-2 \\
        p 
    \end{array}\right)~.\label{eq:prop_dof_massless}
\end{align}
Note that this relation is true for $0 \leq p \leq d-2$. From this result, one can see that a $p$-form and a $(d-p-2)$-form massless gauge theory have the same number of propagating degrees of freedom.

For the case $p=d-1$, the massless gauge theory is topological. Since the field strength tensor of a $(d-1)$-form is a $d$-form, the Hodge dual is a scalar $f\equiv\star \mathrm{d} A_{d-1}$. The equation of motion is $\mathrm{d} f = 0$ which implies that $f$ is constant, showcasing that the theory is non-dynamical, determined up to boundary conditions. The energy-momentum tensor of a massless $(d-1)$-form is $T_{\mu\nu}[A_{d-1}] = \frac{1}{2}g_{\mu\nu}f^2$, so it contributes as a constant energy density in the bulk.

\section{Duality of massive $p$-forms using a topological $B\wedge F$ term}
\label{sec:dual_classic_massive}

Na\"ively, a mass term of the form $\langle A,A\rangle$ seems to be prohibited by the gauge redundancy of a $p$-form theory. There are several ways to overcome this, all of which use some additional degree of freedom to compensate for this violation. In the Standard Model, the Higgs mechanism provides a mass term for the $W^\pm$ and $Z^0$ bosons in the vacuum, contributing scalar Goldstone bosons to act as longitudinal degrees of freedom. In the Stueckelberg theory, these longitudinal degrees of freedom are isolated and the action takes the form
\begin{align}
    S[A,\Omega]&=S_p[A]+\frac{m^2}{2}\int \Big(A-\frac{1}{m}\mathrm{d}\Omega\Big)\wedge \star \Big(A-\frac{1}{m}\mathrm{d}\Omega\Big)~,%
\end{align}
for a $(p-1)$-form Stueckelberg field $\Omega$, which transforms as $\Omega\rightarrow \Omega+m\omega$ while $A\rightarrow A+\mathrm{d}\omega$, thus preserving the gauge invariance of the theory.

To explore the duality between massive theories, we introduce a mass in a different way. Consider a theory with a $p$-form $A$ and a $q$-form $B$, with field strengths $F=\mathrm{d}A$ and $H=\mathrm{d}B$ respectively. If $q=d-p-1$, then we may introduce a topological term coupling the theories, of the form \cite{cremmer_spontaneous_1974,aurilia_u1_1980,allen_topological_1991}
\begin{align}
    im\int B\wedge F=(-1)^{(d-p)(p+1)}im\int A\wedge H+(-1)^{d-p-1}im\int \mathrm{d}(B\wedge A)~,\label{eq:BF coupling}
\end{align}
where $m$ is a parameter with dimensions of mass controlling the coupling strength between the two fields. In the Euclidean theory the term is imaginary, much like the topological $\theta$-term. When we integrate out one of the forms, the resulting theory depends only on $m^2$, so the difference in signs on either side can be absorbed. We will see below that the boundary term vanishes on-shell. The coupling thus behaves symmetrically between the $p$- and $q$-forms. 

The full action of this `$BF$ theory' is then
\begin{align}
    S_{BF}[A,B]=S_p[A] + S_q[B] + im\int B\wedge F~, \label{eq:S_AB_massive}
\end{align}
where $S_p[A]$ is the action for a massless $p$-form as defined in Eq.~\eqref{eq:S[A]-forms}. The theory possesses two independent gauge redundancies of the form $A \rightarrow A+ \mathrm{d}\omega$ and $B \rightarrow B+ \mathrm{d}\theta$. The action is completely gauge invariant under gauge transformations of $A$, while gauge transformations of $B$ leaves the action invariant up to a boundary term
\begin{align}
    S_{BF}[A,B]\rightarrow S_{BF}[A,B]+ im\int\mathrm{d}\left(\theta\wedge F\right)~.
\end{align}
This boundary term is zero if we impose `small' gauge transformations vanishing at the boundary, which represent the true gauge redundancy of the theory. Thus, the action is completely invariant under both gauge transformations of $A$ and $B$.

Varying the action with respect to $A$ gives
\begin{align}
        \delta_A S_{BF}[A,B] = (-1)^{p+1}\int\delta_A A\wedge\left(\mathrm{d} \star F + (-1)^{q(p+1)}im\mathrm{d} B\right) + \int\mathrm{d}\left[\delta_AA\wedge\left(\star F + (-1)^{q(p+1)}imB\right)\right]~, \label{eq:dA_S_AB}
\end{align}
whilst varying with respect to $B$ gives
\begin{align}
        \delta_B S_{BF}[A,B] = (-1)^{q+1}\int\delta_B B\wedge\left(\mathrm{d} \star H + (-1)^{q+1}imF\right) + \int\mathrm{d}\left(\delta_BB\wedge\star H\right)~. \label{eq:dB_S_AB}
\end{align}
Thus, the equations of motion of this theory are
\begin{align}
    \mathrm{d}\star F + (-1)^{q(p+1)}imH= 0~, \quad \mathrm{d} \star H + (-1)^{q+1}imF = 0~, \label{eq:eom_mass}
\end{align}
supplemented by the boundary conditions $\delta_A A|_{\mathrm{bdry}} = \delta_B B|_{\mathrm{bdry}} = 0$ asymptotically (we discuss the possibility of non-trivial gauge patching below).

The second equation of motion can be factorised, locally, as $\mathrm{d}\left(\star H + (-1)^{q+1}m A\right) = 0$. Then in each local patch there is a $(p-1)$-form $\Omega$ such that
\begin{align}
    \star H = (-1)^{q}im\left(A - \frac{1}{m}\mathrm{d}\Omega\right) \implies H = (-1)^{q + p(q+1)}im\star \left(A - \frac{1}{m}\mathrm{d}\Omega\right)~, \label{eq:eom_HA_dual}
\end{align}
We see that $\Omega$ acts as a Stueckelberg field for the $p$-form. As before, we assign it the gauge redundancy $\Omega \rightarrow \Omega+m\omega$ to ensure the theory remains gauge invariant and $H$ is globally smooth. Substituting this solution into the equation of motion of the $p$-form in Eq.~\eqref{eq:eom_mass}, we get the equation of motion of a massive $p$-form
\begin{align}
    \mathrm{d}\star F + (-1)^{p+1}m^2\star\left(A - \frac{1}{m}\mathrm{d}\Omega\right) = 0~,
\end{align}
which is manifestly gauge invariant.

Another way to see this duality is to integrate out the $q$-form in the action using its equation of motion to obtain
\begin{align}
    S[A,\Omega] = S_p[A] + \frac{m^2}{2}\int\left(A - \frac{1}{m}\mathrm{d}\Omega\right)\wedge\star\left(A - \frac{1}{m}\mathrm{d}\Omega\right) - m \int\mathrm{d}\Omega\wedge\star\left(A - \frac{1}{m}\mathrm{d}\Omega\right)~, \label{eq:S_mass_class}
\end{align}
which is almost the action for a massive $p$-form, with an additional term in the end that requires careful treatment. In fact, this final term is a total derivative by virtue of the Bianchi identity, on the constraint imposed by the equation of motion substituted for $H$, Eq.~\eqref{eq:eom_HA_dual},
\begin{align}
    \mathrm{d}\star\left(A - \frac{1}{m}\mathrm{d}\Omega\right)=(-1)^{q-p(q+1)}\frac{1}{im}\mathrm{d}H=0\,.
\end{align}
It can thus be dropped safely.

Furthermore, we are free to fix the gauge\footnote{There is no obstruction to taking this gauge globally, as in fact $F$ is always in the trivial cohomology class---see below.} $A' = A -\frac{1}{m}\mathrm{d}\Omega$ so that we obtain the action of a massive $p$-form explicitly
\begin{align}
    S^{\mathrm{mass.}}_p[A'] = S_p[A'] + \frac{m^2}{2}\int A'\wedge\star A'~.
    \label{eq:S_mass_gauge_fixed}
\end{align}

The $(p-1)$-form $\Omega$ introduces $\left(\begin{array}{c}
    d-2 \\
    p-1
\end{array}\right)$ extra propagating d.o.f.~(c.f.~Eq.~\eqref{eq:prop_dof_massless} for counting), so the total propagating d.o.f.~in the massive theory i.e.~in $A'$ is
\begin{align}
    \text{Propagating d.o.f.~(massive)} = \left(\begin{array}{c}
    d-2 \\
    p
\end{array}\right) + \left(\begin{array}{c}
    d-2 \\
    p-1
\end{array}\right) = \left(\begin{array}{c}
    d-1 \\
    p
\end{array}\right)~, \label{eq:dof_massive}
\end{align}
which is true for $p\leq d-1$.

We now see the utility of starting with a $BF$ theory. Repeating the work above, one can start from the original action Eq.~\eqref{eq:S_AB_massive} and integrate out the $p$-form instead. In this instance, one then obtain the theory of a massive $q$-form. As both theories are equivalent to the parent $BF$ theory, they are themselves dual. Furthermore, Eq.~\eqref{eq:dof_massive} shows that a massive $p$-form and a $q$-form have the same number of propagating degrees of freedom. Thus, the massive $p$-form gauge theory is classically dual to a $(d-p-1)$-form.

More generally we see that there is a web of classical theories, all of which are equivalent on-shell:
\begin{itemize}
    \item The gauge fixed massive $p$- and $(d-p-1)$-form theories, in Eq.~\eqref{eq:S_mass_gauge_fixed}.
    \item The Stueckelberg $p$- and $(d-p-1)$-form gauge theories, in Eq.~\eqref{eq:S_mass_class}
    \item The $BF$ theory of a $p$- and a $(d-p-1)$-form, in Eq.~\eqref{eq:S_AB_massive}.
\end{itemize}
In fact, the relationship between the $BF$ theory and the Stueckelberg theory is itself implied by the \textit{massless} duality: classically the massless $(d-p-1)$-form $B$ is dual to a massless $d-(d-p-1)-2=(p-1)$-form, which is exactly the Stueckelberg field for the coupled $p$-form.

Unlike, for example, massless $p$-form gauge theory or the Abelian Higgs mechanism (at finite neutral Higgs mass) for 1-forms, the Stueckelberg theory does not support configurations where $F$ is in a non-trivial cohomology class, even on topologically non-trivial backgrounds. Consider for example a $1$-form gauge theory on the background of flat four dimensional spacetime punctured by a 't Hooft line, $\mathbb{R}^4\backslash \mathbb{R}\cong S^2\times \mathbb{R}^2$. The field strength is classified by the 2nd cohomology classes $H^2(S^2\times \mathbb{R}^2)\cong \mathbb{Z}$, which correspond to the discrete magnetic charges supported by the 't Hooft line. If the 1-form gains a mass via the Higgs mechanism, the Meissner effect causes the magnetic field to condense into a flux tube. In the centre of the tube, the Higgs field sits at the origin and the symmetry is unbroken. As the Higgs mass is taken to infinity, the flux tube becomes a defect, and the tension of the tube diverges as $T\sim \log m_h$. This suppresses the charged states\footnote{For a single 't Hooft line the action also diverges because the flux tube is infinitely long. Finite length tubes can be made by inserting an additional 't Hooft line with opposite charge; these charged states are likewise suppressed.}. This limit is equivalent to the 1-form Stueckelberg theory up to the insertion of these defects.

The suppression is similarly complete in the general $p$-form Stueckelberg theory. Notice that the field strength, $F=\mathrm{d}A$, can equally be written as $F=\mathrm{d}(A-\frac{1}{m}\mathrm{d}\Omega)$. While the former suggests that some non-trivial patching may be possible, the latter is the derivative of a gauge invariant combination. This combination must be globally defined, so $F$ is always exact and in the trivial class. 

In the $BF$ theory, on-shell the equations of motion for $B$ tell us that $F\propto \mathrm{d}\star H$, so $F$ is likewise exact. The reciprocal conclusion for $H$ seems threatened by the boundary term in Eq.~\eqref{eq:BF coupling} when varying with respect to $A$---in fact this always results in trivial cohomology for $H$. Suppose that we start by assuming some non-trivial patching in $B$, $B=B^i$ on each patch $U^i$ (overlapping only on their boundaries). The variation of $A$ then receives a boundary term proportional to
\begin{align}
    \sum_i\int_{U^i} \mathrm{d}(\delta_AA\wedge B^i)=\sum_{i<j}\int_{\partial U^i\cap (-\partial U^j)} \delta_A A\wedge (B^i-B^j)
\end{align}
which just fixes $B$ to be smooth across patches, and thus forces $H$ in the trivial cohomology class.

This topological triviality on-shell is essential for the classical duality of these theories, otherwise there would exist topologically non-trivial states in one theory which may have no equivalent in its dual. We will show that while this remains true for the field strength in the quantum $BF$ theory, the same cannot be said for the ghost modes. This is what breaks the duality.

\section{Quantisation of massive $p$-forms}
\label{sec:quantise}

Now that we have demonstrated the equivalence of $p$-forms in the classical limit, we will show that the equivalence does not necessarily hold when one considers quantum corrections. Let us begin with the generating functional of our theory
\begin{align}
    Z = \int\mathcal{D} A\mathcal{D} B\Big|_\mathrm{g.f.}\, e^{-S_{BF}[A,B]} ~, \label{eq:Z_original}
\end{align}
where $S_{BF}[A,B]$ is our Euclidean $BF$ action in Eq.~\eqref{eq:S_AB_massive}, and we mark some gauge fixing prescription to be included. We would like to show once again that this theory describes a massive $p$-form by integrating out the $q$-form $B$ in the path integral. Indeed, we would like to write $Z$ in the form
\begin{align}
    Z = X\int\mathcal{D} A'\,e^{-S_p^\mathrm{mass.}[A']}=Y\int \mathcal{D} B'\,e^{-S_q^\mathrm{mass.}[B']} ~, \label{eq:Z_A_eff_action}
\end{align}
where $S_p^\mathrm{mass}[A']$ is the action for a massive $p$-form, as Eq.~\eqref{eq:S_mass_gauge_fixed}. We will show that $X$ and $Y$ consist of determinants which produce different counterterms breaking the duality between the theories. In fact, these factors will simply correspond to the failure to dualise the form that is integrated out to a longitudinal degree of freedom for the remaining form.

To begin, we will complete the path integral over $B$, which we expect to contain a mass term for $A$.  This path integral is just the generating functional of a free $q$-form gauge theory with a source term $\star J_B \equiv mF$\
\begin{align}
    Z_q[A] \equiv \int\mathcal{D} B\Big|_\mathrm{g.f.}\,\exp\left(-\frac{1}{2}\langle \mathrm{d} B,\mathrm{d} B\rangle - im\int B\wedge  F\right)~. \label{eq:Z_q}
\end{align}
We quantise purely in the sector where $H$ is in the trivial cohomology class; thus we can take $B$ to be global and no boundary terms arise from patching. This is the sector where the classical theory lives, and the only sector where we expect the theories to be dual. In the other sectors, which possess no classical solutions, the non-trivial cohomology of $H$ seems to source the $A$ field in such a way as to break the duality directly---we defer these to later study. We will see that imposing trivial cohomology for $H$ dynamically recovers the trivial topology for $F$ as in the classical theory. This is in contrast to the case of massless $p$-forms, where harmonic field strength contributions have been shown to restore the massless duality in odd dimensions \cite{donnelly_electromagnetic_2017}.

We will assume that the spacetime is Euclidean and geodesically complete, ensuring that the Poincar\'e duality survives \cite{dai_introduction_nodate} and there are no boundary terms for normalisable forms. See Appendix~\ref{app:hodge theory} for some more discussion. %

To solve this path integral, we need to to impose some gauge-fixing procedure on $B$ to avoid overcounting gauge-equivalent paths. The path integral contains an infinite volume contribution from the gauge direction, which we need to factorise out and absorb into the normalisation of $Z_q[A]$. Rather than taking the traditional Faddeev--Popov approach, where ghost counting is perilous (e.g.~\cite{siegel_hidden_1980}), we follow an approach similar to \cite{obukhov_geometrical_1982}. We write the gauge transformations of the $q$-form as $ B[\theta]\equiv B+\mathrm{d} \theta$. Suppose that there exists a functional $G_q[B[\theta]]$ such that
\begin{align}
    1=\int \mathcal{D} \theta\, G_q[B[\theta]]\,\exp\left(-\frac{1}{2}\langle \delta B[\theta],\delta B[\theta]\rangle\right)~,\label{eq:resolution of identity}
\end{align}
where $\delta$ is the codifferential. Inserting this identity into Eq.~\eqref{eq:Z_q} recovers the usual Loren$\delta$z gauge result for the gauge-fixed action
\begin{align}
    Z_q[A]=\int\mathcal{D}\theta\int \mathcal{D} B \,G_q[B]\,\exp\!\left(\frac{1}{2}\langle B,\Box B\rangle - im\int B\wedge F \right)~,
\end{align}
where $\Box=-(\delta \mathrm{d}+\mathrm{d} \delta)$ is the Hodge Laplacian. The remaining integral over $\theta$ is an overall factor of the gauge group volume that may be absorbed into the normalisation of $Z_q[A]$. Here we see that $G_q[B]$ encodes the gauge-fixing condition of the theory. In Appendix~\ref{app:gauge_fix}, we show that $G_q[B]$ is independent of $B$ and that
\begin{align}
    G_q[B] = V_{\mathrm{gauge}}\mathcal{G}_q \equiv V_{\mathrm{gauge}}\prod_{k=1}^q\big(\det\Box_{q-k}\big)^{(-1)^{k+1}(k+1)/2}=V_{\mathrm{gauge}}\det\Box_{q-1}^{+1}\det\Box_{q-2}^{-3/2}\cdots\det\Box_0^{(-1)^{q+1} (q+1)/2}~, \label{eq:G_q_tower_ghosts}
\end{align}
where $V_{\mathrm{gauge}}$ is an infinite gauge volume factor that we can absorb into the normalisation of the path integral and $\mathcal{G}_q$ contains the tower of ghost fields associated with the $q$-form (see, for example, Duff and Van Nieuwenhuizen \cite{duff_quantum_1980}). The subscript on the Hodge Laplacians indicates the degree of the forms on which it acts. Here, the alternating contributions in successive degrees reproduces the competing gauge redundancies seen in the classical counting of degrees of freedom, Eq.~\eqref{eq:prop_dof_massless}. Thus, the properly normalised path integral (with the gauge volume absorbed) is
\begin{align}
   Z_q[A]=\bigg[\prod_{k=1}^q\big(\det\Box_{q-k}\big)^{(-1)^{k+1}(k+1)/2}\bigg]\int \mathcal{D} B \,\exp\left(\frac{1}{2}\langle B,\Box B\rangle  - im\int B\wedge F\right)~. \label{eq:Z_q[A]_gauge_fixed}
\end{align}

Taking $m=0$ recovers the path integral for a massless $q$-form decoupled from $A$:
\begin{align}
    Z_q^0\equiv\mathcal{G}_q\det\Box^{-1/2}_q=\prod_{k=0}^q\big(\det\Box_{q-k}\big)^{(-1)^{k+1}(k+1)/2}~\label{eq:massless q-form path integral},
\end{align}

For non-zero $m$, we now will complete the path integral over $B$, in the $q$-th degree. To complete the square, we want to shift $B\rightarrow B+(-1)^{\eta}im\Box^{-1}\star F$  (except for zero modes---see below) where $\star^2 F=(-1)^\eta F$, $\eta=(d-p-1)(p+1)$, so that
\begin{align}
    \frac{1}{2}\int B\wedge\star \Box B-im\int B\wedge F &\rightarrow \frac{1}{2}\int ( B+(-1)^{\eta}im\Box^{-1}\star F)\wedge \star( \Box B+(-1)^\eta im\star F )\\
    &\hspace{5cm}-im\int (B+(-1)^\eta im\Box^{-1}\star F)\wedge F \nonumber\\
    &=\frac{1}{2}\langle B,\Box B\rangle+\frac{m^2}{2}\langle \Box^{-1}\star F,\star F\rangle ~.
\intertext{Then using the fact that $\Box$ is Hermitian and commutes with $\star$, %
and that the inner product is duality invariant, we can write this as}
    &=\frac{1}{2}\langle B,\Box B\rangle+\frac{m^2}{2}\langle F, \Box^{-1} F\rangle ~.
\end{align}
Now we can freely integrate out $B$, leaving
\begin{align}
    Z_q[A]=\mathcal{G}_q\det\Box_{q}^{-1/2}\exp\!\Big(\frac{m^2}{2}\langle F,\Box^{-1}F\rangle\Big)=Z_q^0\exp\!\Big(\frac{m^2}{2}\langle F,\Box^{-1}F\rangle\Big)~,
\end{align}
where the final determinant once again combines with the ghosts to leave a massless $q$-form path integral $Z^0_q$ as above, Eq.~\eqref{eq:massless q-form path integral}.

Where $F$ is harmonic, the change of variable is badly defined. To regulate this, we can insert a small fictitious mass $\mu$ and perform the shift $\Box\rightarrow \Box-\mu^2$. The resulting path integral is
\begin{align}
    Z_q[A]=\mathcal{G}_q\lim_{\mu\rightarrow 0}\det\!\big(\Box_{q}-\mu^2\big)^{-1/2}\exp\!\Big(\frac{m^2}{2}\langle F,(\Box-\mu^2)^{-1}F\rangle\Big)~.
\end{align}
However, on any zero mode $F_0$ we have that $\Box F_0=0$, so for all $\mu^2>0$
\begin{align}
    \exp\!\Big(\frac{m^2}{2}\langle F_0,(\Box-\mu^2)^{-1}F_0\rangle\Big)=\exp\!\Big(-\frac{m^2}{2\mu^2}\langle F_0, F_0\rangle\Big)\,.
\end{align}
Along with the usual kinetic term for $F_0$, this produces a Gaussian path integral of width $\sim\mu/m$. The limit $\mu\rightarrow 0$ then leads to a delta function\footnote{Compare this to the limit of the Landau gauge $\xi\rightarrow 0$ where the gauge fixing term $\frac{1}{2\xi}(\partial_\mu A^\mu )^2$ forces $\partial_\mu A^\mu=0$ in QED.} around $F_0=0$, killing the non-trivial cohomology classes of $F$. Alternatively, notice that the harmonic modes of $B$ have vanishing kinetic term, but still couple to $F_0$, acting as a Lagrange multiplier forcing $F_0=0$. 
This matches the suppression of charges in non-trivial classes seen in the classical theory. 

More generally, we might consider the presence of an additional $q$-form source $J_B$ for the $B$ field, which is assumed to be conserved, $\mathrm{d}\star J_B$, to ensure gauge invariance. In this case, the coupling term $\langle B,J_B\rangle$ in the action shifts the resulting contribution to the effective action, leaving
\begin{align}
    \frac{m^2}{2}\langle (F+\star J_B),\Box^{-1}(F+\star J_B)\rangle~.
\end{align}
On the zero modes, this shifts the resulting delta function such that $F$ lies in the same cohomology class as $-(\star J_B)$. The remaining non-zero modes of $\star J_B$ then couple to the transverse modes of $A$.

On non-zero modes, the $\langle F,\Box^{-1}F\rangle$ term is a gauge invariant mass term for the $A$ field. Its behaviour can be made clearer by integrating by parts
\begin{align}
    -\frac{m^2}{2}\langle F,\Box^{-1}F\rangle=-\frac{m^2}{2}\langle \mathrm{d} A,\Box^{-1}\mathrm{d} A\rangle=+\frac{m^2}{2}\Big(\langle A,(1+\Box^{-1}\mathrm{d}\delta) A\rangle\Big)=\frac{m^2}{2}\langle A^T,A^T\rangle~, \label{eq:gauge invariant}
\end{align}
where we have Hodge decomposed $A=A^T+\mathrm{d}\alpha$ into transverse ($\delta A^T=0$) and longitudinal modes, as
\begin{align}
    -\delta \Box^{-1}\mathrm{d} = -\Box^{-1}\delta\mathrm{d} = -\Box^{-1}(-\Box-\mathrm{d}\delta)=1+ \Box^{-1}\mathrm{d}\delta\equiv \Pi_T
\end{align}
is exactly the projection operator $\Pi_T$ onto transverse modes. Indeed, we can verify this by directly inserting the decomposition,
\begin{align}
    \Pi_T A=(1+ \Box^{-1}\mathrm{d}\delta)(A^T+\mathrm{d} \alpha)=A^T+\mathrm{d}\alpha+\Box^{-1}(\mathrm{d}\delta )\mathrm{d} \alpha=A^T+\mathrm{d}\alpha+\Box^{-1}(-\Box) \mathrm{d} \alpha=A^T\,.
\end{align}

Thus we see that integrating out the $B$ field produces
\begin{enumerate}
    \item[(1)] A tower of massless degrees of freedom $Z^0_q$ for the $q$-form theory.

    \item[(2)] A gauge invariant mass term for the transverse modes of the $p$-form $A$.

    \item[(3)] A constraint restricting $F$ to the trivial cohomology class.
\end{enumerate}
The full path integral is
\begin{align}
    Z&=Z^0_q\int \mathcal{D} A\Big|_\mathrm{g.f.}\,\exp\!\left(-\frac{1}{2}\langle F, F\rangle-\frac{m^2}{2}\langle A^T,A^T\rangle\right)~. \label{eq:Z B integrated out}
\end{align}
The integral over $A$ is gauge invariant, but is non-local (owing to the projector containing a $\Box^{-1}$ term) and contains only the transverse modes of $A$.

To produce a proper path integral for a massive $p$-form, we want to explicitly restore the longitudinal modes. We do this by multiplying and dividing Eq. \eqref{eq:Z B integrated out} by the path integral for a massless $(p-1)$-form $\pi$. This results in 
\begin{align}
    Z=\frac{Z^0_q}{Z_{p-1}^0}\int \mathcal{D} A\mathcal{D}\pi\Big|_\mathrm{g.f.}\,\exp\!\left(-\frac{1}{2}\langle F, F\rangle-\frac{m^2}{2}\langle A^T,  A^T\rangle -\frac{1}{2}\langle\mathrm{d}\pi, \mathrm{d}\pi\rangle\right) ~. \label{eq:Z massive p-form}
\end{align}
The factor $Z_q^0/Z_{p-1}^0$ is simply the ratio of classically dual \textit{massless} theories, where the $q$-form $B$ has been dualised to a $(p-1)$-form Stueckelberg field as in the classical case. In the quantum theory the duality is broken \cite{duff_quantum_1980}, so this factor becomes non-trivial. Finally if we define the gauge invariant potential $A'=A^T+\frac{1}{m}\mathrm{d}\pi$, the path integral in Eq. \eqref{eq:Z massive p-form} can be succintly written as
\begin{align}
    Z=\frac{Z^0_q}{Z_{p-1}^0}\int \mathcal{D} A'\,\exp\!\left(-\frac{1}{2}\langle F, F\rangle-\frac{m^2}{2}\langle A',  A'\rangle\right)=\frac{Z_q^0}{Z_{p-1}^0}Z^m_{p}~,
\end{align}
where
\begin{align}
    Z_p^m\equiv\int \mathcal{D} A'\,\exp\!\left(-\frac{1}{2}\langle F, F\rangle-\frac{m^2}{2}\langle A',  A'\rangle\right)\label{eq:gauge fixed path integral}~,
\end{align}
is the path integral for a massive $p$-form. We see that a quantum $BF$ theory equivalent to a massive $p$-form theory, up to the failure of the $B$ field to dualise to the massless longitudinal modes. 

On the other hand, we could have equally integrated out $A$ instead, giving
\begin{align}
    Z=\frac{Z^0_p}{Z_{q-1}^0}\int \mathcal{D} B'\,\exp\!\left(-\frac{1}{2}\langle H,H\rangle-\frac{m^2}{2}\langle B', B'\rangle\right)=\frac{Z_p^0}{Z_{q-1}^0}Z^m_{q}~.
\end{align}
Now we see that the (non)equivalence of the quantum theories is given exactly by the (non)equivalence of the dualised longitudinal modes, ${Z_q^0}/{Z_{p-1}^0}\neq {Z_p^0}/{Z_{q-1}^0}$. Indeed, we see that the ratio of path integrals is \textit{independent of mass!}\footnote{Note however, that this isn't to say that the (non)equivalence is the same as that of massless theories themselves, as in this case the appropriate $p$ and $q$ are aligned differently, $q\overset{!}{=}d-p-2$. Furthermore, the massless limit results in a massless $p$-form accompanied by a decoupled longitudinal $(p-1)$-form.} 
\begin{align}
     \frac{Z_p^m}{Z_q^m}=\frac{Z^0_p Z_{p-1}^0}{Z^0_q Z_{q-1}^0}~. \label{eq:quotient of theories}
\end{align}

\section{Zero modes and counterterms}

As discussed by Duff and Van Nieuwenhuizen \cite{duff_quantum_1980}, zero modes of $\det\Box_k^{n/2}$ produce divergences which must be absorbed by a local counterterm $\Delta \mathcal{L}=-\frac{\sqrt{g}}{\varepsilon}n b_k$, along with appropriate boundary contributions---regularised in $d+\varepsilon$ dimensions at one loop---such that the action counts the corresponding \textit{normalisable} %
Betti number %
\begin{align}
    B_k=\int \mathrm{d}^d x\sqrt{g}\,b_k+\text{(boundary terms)}~.
\end{align}
By definition, this is the number of normalisable zero modes of $\Box_k$, i.e.~normalisable harmonic $k$-forms (see Appendix~\ref{app:hodge theory}).
After integration, the counterterms in the action are $\Delta S=-\frac{1}{\varepsilon}n B_k$.
Using the general form for $Z_p^0$, the total counterterm is
\begin{align}
    \Delta S_p= \frac{1}{\varepsilon}(B_p-2B_{p-1}+\cdots +(-1)^{p}(p+1)B_0)~.
\end{align}
The counterterm for the numerator of Eq.~\eqref{eq:quotient of theories} is then
\begin{align}
    \Delta S_p+\Delta S_{p-1}&=\frac{1}{\varepsilon}\Big[ (B_p-2B_{p-1}+\cdots +(-1)^{p}(p+1)B_0)+(B_{p-1}-2B_{p-2}+\cdots +(-1)^{p-1}p B_0)\Big] \\
    &=\frac{1}{\varepsilon}\Big[B_p-B_{p-1}+\cdots +(-1)^p B_0\Big]~,
\end{align}
and likewise the denominator
\begin{align}
    \Delta S_q+\Delta S_{q-1}&=\frac{1}{\varepsilon}\Big[B_q-B_{q-1}+\cdots +(-1)^q B_0\Big]~.
\end{align}
Thus the counterterms in $p$-form and $q$-form theories differ by
\begin{align}
    \Delta S^\mathrm{mass.}_{p\leftrightarrow q}&=\Delta S_p+\Delta S_{p-1}-\Delta S_q-\Delta S_{q-1} \\
    &=\frac{1}{\varepsilon}\Big[(B_p-B_{p-1}+\cdots +(-1)^p B_0)-(B_{d-p-1}-B_{d-p-2}+\cdots +(-1)^{d-p-1} B_0)\Big]~,
\intertext{
then using the symmetry of normalisable Betti numbers on geodesically complete manifolds}
    &=\frac{1}{\varepsilon}\Big[(B_p-B_{p-1}+\cdots +(-1)^p B_0)-(B_{p+1}-B_{p+2}+\cdots +(-1)^{d-p-1} B_d)\Big]\\
    &=\frac{(-1)^p}{\varepsilon}\Big[B_0-B_1+B_2-\cdots +(-1)^d B_d\Big]=\frac{(-1)^{p}}{\varepsilon}\chi^{(2)}~,
\end{align}
where $\chi^{(2)}=\chi$ is the Euler characteristic of a compact space. More care must be taken in the case of non-compact spaces, where the $\chi^{(2)}$ may only equal the full Euler characteristic up to boundary corrections---here the alternating sum of \textit{normalisable} Betti numbers gives the right expression. In either case, the duality between the theories is broken by a topological invariant. %

Explicit local Lagrangian counterterms may be computed in terms of the Riemann tensor and its traces, using, for example, heat kernel methods \cite{christensen_new_1979}.

This result is directly comparable to the breaking of the duality between massless $p$- and $(q=d-p-2)$-form theories studied by Duff and Van Nieuwenhuizen in $d=4$ \cite{duff_quantum_1980}, who concluded that the duality is broken by counterterms proportional to $\chi$. Indeed, our result can be seen as a direct consequence, as the breaking is seen to come only from the inequivalent dualisations of the integrated fields to the massless longitudinal degree of freedom of their partner.

\section{Examples of quantum non-equivalence}
\label{sec:examples}

We will now give some explicit examples in $4$ dimension. In this case, the pairing of theories is
\begin{align}
    \text{$0$-forms}\quad &\longleftrightarrow \quad \text{$3$-forms} \\
    \text{$1$-forms}\quad &\longleftrightarrow \quad \text{$2$-forms}~,
\end{align}
for which we have
\begin{align}
    \Delta S^\mathrm{mass.}_{0\leftrightarrow 3}&=\frac{+1}{\varepsilon}(B_0-B_1+B_2-B_3+B_4) \\
    \Delta S^\mathrm{mass.}_{1\leftrightarrow 2}&=\frac{-1}{\varepsilon}(B_0-B_1+B_2-B_3+B_4)~.
\end{align}
The local counterterms are proportional to
\begin{align}
    b_0-b_1+b_2-b_3+b_4=\frac{1}{32\pi^2}(R_{\mu\nu\rho\sigma}R^{\mu\nu\rho\sigma}-4R_{\mu\nu}R^{\mu\nu} +R^2)~,
\end{align}
which is a total divergence, and integrates to the Euler characteristic (up to boundary corrections) by the Gauss--Bonnet theorem \cite{duff_quantum_1980}.

In flat Euclidean space, all the normalisable Betti numbers vanish, so the theories are dual. We will look at three  particular non-trivial cases in detail: the Euclidean Schwarzschild spacetime, the Euclidean de Sitter spacetime, and the Eguchi--Hansen instanton.

\subsection{Euclidean Schwarzschild spacetime}

The Euclidean Schwarzschild spacetime $M$ describes the thermal Hawking radiation of non-rotating neutral black holes \cite{gibbons_action_1977}. It is described by the metric
\begin{align}
    \mathrm{d} s^2=\Big(1-\frac{r_\mathrm{h}}{r}\Big)\mathrm{d} \tau^2+\Big(1-\frac{r_\mathrm{h}}{r}\Big)^{-1}\mathrm{d} r^2 + r^2(\mathrm{d} \theta^2+\sin^2\theta\,\mathrm{d} \phi^2)\,,
\end{align}
covering the whole manifold except the 2-sphere at $r=r_\mathrm{h}$, the event horizon.
To make the spacetime everywhere regular (so the spacetime can be interpreted as an smooth continuation of its Lorentzian counterpart) the imaginary time coordinate $\tau$ coordinate must be made periodic. The resulting periodicity is exactly the reciprocal of the Hawking temperature, $4\pi r_\mathrm{h}=1/T_\mathrm{H}$, and loops in $\tau$ shrink to a point at $r=r_\mathrm{h}$, excising the interior of the horizon from the manifold.

This spacetime is homotopic to $\mathbb{R}^2\times S^2$, where $r$ and $\tau$ act as the radial and polar coordinates of $\mathbb{R}^2$, respectively, and $S^2$ is spanned by the usual angular coordinates, $\theta$ and $\phi$. Because $M$ is asymptotically locally flat, normalisable $2$-forms on $M$ are equivalent to $2$-forms on $M$ compactified with a single point at infinity \cite{hausel_hodge_2003}. This changes the relevant topology to $M\cup \{\infty\}\cong S^2\times S^2$. As described in \cite{kobakhidze_electromagnetic_2026}, this topology supports harmonic 2-forms $\omega_{(n,k)}$ wrapped around each 2-sphere, labelled by integer electric and magnetic charges, $(n,k)\in\mathbb{Z}\oplus \mathbb{Z}$
\begin{align}
    \omega_{(n,k)} = \frac{n}{2r^2}\mathrm{d} \tau \wedge \mathrm{d} r + \frac{k}{2}\sin\theta\,\mathrm{d}\theta\wedge \mathrm{d}\phi~.
\end{align}
There are no other normalisable harmonic $p$-forms on $M$ \cite{etesi_geometric_2001}, so
\begin{align}
    B_k=\begin{cases}
        2\,, & k=2 \\
        0\,, & \text{else}
    \end{cases}\quad.
\end{align}
This reproduces the Euler characteristic for the non-compactified spacetime, $\chi=\sum_k (-1)^kB_k=2$ (c.f.~\cite{hawking_gravitational_1977}) with no boundary correction.
Thus
\begin{align}
    \Delta S^\mathrm{mass.}_{0\leftrightarrow 3}&=\frac{2}{\varepsilon}\\
    \Delta S^\mathrm{mass.}_{1\leftrightarrow 2}&=-\frac{2}{\varepsilon}~,
\end{align}
so both dualities are broken.

\subsection{Euclidean de Sitter}
Because the Lorentzian de Sitter space lacks a global timelike Killing vector, an ordinary theory of scattering amplitudes is badly defined \cite{dyson_disturbing_2002,witten_quantum_2001}; on the other hand, an observer seeing a `static patch' bounded by cosmological event horizons admits a thermal description \cite{gibbons_action_1977}. Much like the Schwarzschild spacetime, these thermal properties of the de Sitter spacetime may be described by the analytic continuation of the static patch to imaginary time. The resulting manifold is simply the $d$-sphere, $S^d$. The Betti numbers for the $d$-sphere are
\begin{align}
    B_k=\begin{cases}
        1\,, & k=0,d\\ 0\,, & \text{else}
    \end{cases}\quad,
\end{align}
the only harmonic forms being the constant 0-forms and multiples of the volume form. Thus the difference of counterterms in four dimensions are
\begin{align}
    \Delta S^\mathrm{mass.}_{0\leftrightarrow 3}&=\frac{2}{\varepsilon}\\
    \Delta S^\mathrm{mass.}_{1\leftrightarrow 2}&=-\frac{2}{\varepsilon}~,
\end{align}
so both dualities are also broken in this spacetime. This shows that the classical duality of massive $p$-forms cannot be blindly applied in models of inflation, being broken at least in the thermal spectrum. This may impact the study of the reheating stage at the end of $p$-inflation models \cite{de_felice_reheating_2012,mulryne_three-form_2012}. We remark that in odd dimensions, the top and bottom form contributions cancel, and the duality is preserved.

\subsection{Eguchi--Hanson instantons}

The Eguchi--Hanson instantons \cite{eguchi_asymptotically_1978,eguchi_self-dual_1979} are the asymptotically locally Euclidean spacetimes with metric
\begin{align}
    \mathrm{d} s^2=\Big(1-\frac{a^4}{r^4}\Big)^{-1}\mathrm{d} r^2+r^2 (\sigma_x^2+\sigma_y^2) + r^2 \Big( 1-\frac{a^4}{r^4}\Big)\sigma_z^2\,,
\end{align}
where $a$ is the size of the instanton and $\sigma_i$ are $\mathrm{SU}(2)$ Cartan forms, satisfying the structure equation $\mathrm{d} \sigma_i=\varepsilon_{ijk}\,\mathrm{d}\sigma_j\wedge\mathrm{d}\sigma_k$. These spacetimes are proposed to describe tunnelling in quantum gravity in analogy to Yang--Mills instantons, leading to gravitational $\theta$-vacua \cite{dvali_hint_2024}.

The unique $L^2$ harmonic forms on the Eguchi--Hansen spacetime are the constant multiples of the anti-self-dual 2-form \cite{eguchi_self-dual_1979}
\begin{align}
    \omega%
    =\frac{1}{r^2}\sigma_x\wedge\sigma_y-\frac{1}{r^3}\mathrm{d}r\wedge \sigma_z \,,
\end{align}
which is locally the exterior derivative of $\sigma_z/2r^2$. Thus the normalisable Betti numbers are
\begin{align}
    B_k=\begin{cases}
        1\,, & k=2 \\
        0\,, & \text{else}
    \end{cases}\quad.
\end{align}
Note that the Euler characteristic receives a boundary correction, such that $\chi=2$. Thus
\begin{align}
    \Delta S^\mathrm{mass.}_{0\leftrightarrow 3}&=\frac{1}{\varepsilon}\\
    \Delta S^\mathrm{mass.}_{1\leftrightarrow 2}&=-\frac{1}{\varepsilon}~,
\end{align}
so both dualities are broken yet again.

This case is of particular interest, even for observers in flat space; if these spacetimes do represent instantons in quantum gravity as expected, then these differing topological numbers demonstrate that the duality between massive $p$-form theories is broken by quantum gravity even in the vacuum.   
\section{Conclusion}
\label{sec:conc}

In this paper we have demonstrated that the classical duality between massive $p$- and $(d-p-1)$-form gauge theories is broken by quantum corrections on topologically non-trivial spacetime backgrounds. The breaking arises from the mismatch of zero modes, which demand counterterms differing by a topological invariant. This invariant is the Euler characteristic of the spacetime, up to possible boundary corrections on non-compact manifolds.

The challenge of quantising a $p$-form gauge theory comes from the treatment of the gauge invariance, both in producing an appropriately gauge fixed path integral and gauge invariant mass term. For the former, we exhibit a novel gauge fixing functional. We achieve the latter with using the $BF$ theory, where the manifestly gauge invariant topological coupling of a $p$- and $(d-p-1)$-form produces a mass term for one of the forms when the other is integrated out. In this formulation the non-equivalence is related to the quantum breaking of the massless $p$-form duality through the dualisation of the integrated-out form to a Stueckelberg field for its partner.

Considering specific examples in four dimensions, we have shown that the duality is broken in the thermal theories on Schwarzschild and de Sitter spaces. We have also shown that the duality is broken on the Eguchi--Hansen spacetime, which is thought to describe instantons in quantum gravity. These results suggest that the duality between massive theories may be broken even in the gravitational vacuum.

The essential topological structure of $p$-form gauge theories is that of de Rham cohomology. We have emphasised that, in both the classical and quantum cases, it is essential that the field strengths for dual massive $p$-form theories live in the trivial cohomology classes. In contrast to the massless theory, where the duality connects fields strengths in dual degrees, $\star F=G$, fields strengths for the massive theories in non-trivial cohomology classes would immediately violate the duality on backgrounds with non-isomorphic cohomology groups in the $p$-th and $(p-d-1)$-th degrees. In the classical theory this is enforced by the dynamics on-shell automatically. In the quantum $BF$ theory the situations is less clear: choosing to quantise one field in the trivial sector automatically ensures that the other is trivial as well; on the other hand, the (purely quantum) theories in the non-trivial sectors require further study.

Unlike the field strengths, the cohomology classes of the $p$-form potential and ghost fields may be non-trivial. For each dimension of the cohomology group there is a corresponding zero mode of the Hodge Laplacian. These zero modes have zero action, so they contribute divergences that must be renormalised. The resulting counterterms break the duality in the quantum theory. This result, in essence, is a consequence of the topology of the gauge structure.

Our results are applicable to geodesically complete spacetimes, where the space of normalisable zero modes is sufficiently well-behaved. On spacetimes with a non-asymptotic boundary, boundary corrections will likely become more important; here a treatment of edge modes (see, for example, \cite{ball_dynamical_2024}) would be necessary.

\section{Acknowledgements}

The authors would like to thank Archil Kobakhidze for his invaluable guidance and support throughout the development of this work. They also thank Otari Sakhelashvili for his keen insights regarding $p$-form theories, especially those which helped clarify the behaviour of the gauge invariant mass operator. The authors also acknowledge support from the University of Sydney Grant-in-Aid and PRSS travel grants.

\appendix
\section{Normalisation and components of $p$-forms}
\label{app:note}
\renewcommand{\theequation}{A-\arabic{equation}}
 \setcounter{equation}{0}

In this appendix we fix conventions relating differential forms and their components in a basis, as well as the normalisation of the field strength and kinetic terms of $p$-form gauge theories.

Let $\omega$ and $\chi$ be $p$-forms over a Riemannian manifold. Given a basis, a form and its components are related by
\begin{align}
    \omega=\frac{1}{p!}\omega_{\mu_1\cdots\mu_{p}}\mathrm{d}x^{\mu_1}\wedge \cdots \wedge \mathrm{d}x^{\mu_p}~,
\end{align}
where $\omega_{\mu_1\cdots\mu_{p}}$ is totally antisymmetric. We work in Euclidean signature, in $d$ dimensions. Given a metric $g_{\mu\nu}$ with determinant $g$, the Hodge dual $\star$ acts by
\begin{align}
    \star \omega=\frac{1}{(d-p)!}(\star \omega)_{\mu_{p+1}\cdots \mu_{d}}\mathrm{d}x^{\mu_{p+1}}\wedge \cdots \wedge \mathrm{d}x^{\mu_d}=\frac{1}{p!(d-p)!}\varepsilon_{\mu_1\cdots\mu_d}\omega^{\mu_{1}\cdots \mu_p}\mathrm{d}x^{\mu_{p+1}}\wedge \cdots \wedge \mathrm{d}x^{\mu_d}\,,
\end{align}
with components
\begin{align}
    (\star \omega)_{\mu_{p+1}\cdots \mu_{d}}=\frac{1}{p!}\varepsilon_{\mu_1\cdots\mu_d}\omega^{\mu_{1}\cdots \mu_p}~,
\end{align}
where $\varepsilon_{\mu_1\cdots \mu_d}$ is the covariant Levi-Civita tensor, with $\varepsilon_{12\cdots d}=\sqrt{g}$. The Hodge dual squares to $\star^2\omega=(-1)^{p(d-p)}\omega$ acting on $p$-forms.%

The Hodge dual defines a natural inner product on $p$-forms over a manifold
\begin{align}
    \langle\omega,\chi\rangle\equiv\int \omega\wedge\star \chi&=\frac{1}{p!}\int \omega_{\mu_1\cdots\mu_p}\chi^{\mu_1\cdots \mu_p}\sqrt{g}\,\mathrm{d}^dx~,
\end{align}
where $\mathrm{d}^dx\equiv\mathrm{d}x^1\wedge \cdots \wedge \mathrm{d}x^d$ is the volume form. %
The inner product is symmetric and duality invariant, $\langle\star \omega,\star\chi\rangle=\langle \omega,\chi\rangle$. A $p$-form $\omega$ is said to $L^2$ normalisable if $\langle\omega,\omega\rangle<\infty$.

The exterior derivative $\mathrm{d}$ maps $p$-forms to $(p+1)$-forms, with
\begin{align}
    \mathrm{d}\omega=\frac{1}{p!}\partial_{\mu_0}\omega_{\mu_1\cdots\mu_{p}}\mathrm{d}x^{\mu_0}\wedge \cdots \wedge\mathrm{d}x^{\mu_{p}}=\frac{1}{p!}\partial_{[\mu_0}\omega_{\mu_1\cdots\mu_{p}]}\mathrm{d}x^{\mu_0}\wedge \cdots \wedge\mathrm{d}x^{\mu_{p}}~,
\end{align}
where $[\,\cdots]$ denotes the \textit{normalised} antisymmetrisation of indexes.
The components are then
\begin{align}
    (\mathrm{d}\omega)_{\mu_0\cdots \mu_{p}}=(p+1)\partial_{[\mu_0}\omega_{\mu_2\cdots\mu_{p}]}~.
\end{align}

Most importantly, this normalises the field strength for a $p$-form
\begin{align}
    F=\mathrm{d}A=\frac{1}{p!}\partial_{[\mu_0} A_{\mu_2\cdots \mu_{p}]}\mathrm{d}x^{\mu_0}\wedge \cdots \wedge\mathrm{d}x^{\mu_{p}}\,,\qquad F_{\mu_0\cdots \mu_{p}}=(p+1)\partial_{[\mu_0}A_{\mu_1\cdots \mu_{p}]}~.
\end{align}
The kinetic term in the action is then normalised
\begin{align}
    S_p[A]&=\frac{1}{2}\int F\wedge \star F=\frac{1}{2(p+1)!}\int F_{\mu_0\cdots \mu_{p}}F^{\mu_0\cdots \mu_{p}}\sqrt{g}\,\mathrm{d}^dx~,
\intertext{or in terms of the potential}
    &=\frac{(p+1)}{2p!}\int \partial_{[\mu_0}A_{\mu_1\cdots \mu_{p}]}\partial^{[\mu_0}A^{\mu_1\cdots \mu_{p}]}\sqrt{g}\,\mathrm{d}^dx~.
\end{align}
By expanding the antisymmetrisation, one can check that this is the canonical normalisation for each independent degree of freedom.

The codifferential $\delta$ is defined as the adjoint of $\mathrm{d}$ such that $\langle\mathrm{d}\omega,\chi \rangle=\langle\omega,\delta\chi\rangle$ up to boundary terms. Using the definition of the inner product and Stokes theorem, we find that $\delta \omega=(-1)^k\star^{-1}\mathrm{d}\star \omega=(-1)^{(d-k-1)(k+1)+k}\star \mathrm{d}\star\omega$ when acting on a $p$-form. 

We denote by $\Box\equiv-(\delta\mathrm{d}+\mathrm{d}\delta)$ the Hodge Laplacian (a.k.a.~the Laplace--de Rham operator). This sign is opposite to that of much of the mathematical literature, but aligns with the usual Laplacian on scalars. $\Box$ is essentially self-adjoint for normalisable $p$-forms on complete manifolds \cite{chernoff_essential_1973} and negative semi-definite, i.e.~has only negative and zero eigenvalues. When necessary, we mark the operator with a subscript to indicate its restriction to a particular degree, such as when writing its determinant.

\section{Derivation of gauge-fixing functional}
\label{app:gauge_fix}

In this section we provide a detailed derivation of the gauge-fixing method performed in Sec.~\ref{sec:quantise} when quantising the massive $p$-form gauge theory. Firstly, we show that the gauge-fixing functional $G_q[B[\theta_{q-1}]]$ defined in Eq.~\eqref{eq:resolution of identity} is independent of $B$. 

Suppose we have two $q$-forms $B$ and $\tilde{B}$, with Hodge decompositions $B\equiv \mathrm{d}\alpha + \delta \beta + \gamma$ and $\tilde{B} = \mathrm{d}\tilde{\alpha} + \delta\tilde{\beta} + \tilde{\gamma}$ where $\gamma$, $\tilde{\gamma}$ are harmonic forms. The gauge-transformed forms under this decomposition are simply $B[\theta_{q-1}] = \mathrm{d} (\alpha + \theta_{q-1}) + \delta\beta + \gamma$ and similar for $\tilde{B}[\theta_{q-1}]$. The identity Eq.~\eqref{eq:resolution of identity} in terms of $B$ can be written as
\begin{align}
    1=\int \mathcal{D} \theta_{q-1} \,G_q[B[\theta_{q-1}]]\,\exp\!\left(-\frac{1}{2}\langle \delta \mathrm{d}(\alpha + \theta_{q-1}),\delta\mathrm{d}(\alpha + \theta_{q-1})\rangle\right)~.
\end{align}
We then perform a change of variables $\theta_{q-1} \to \theta_{q-1}' = \alpha + \theta_{q-1}$, leaving the functional measure invariant, giving 
\begin{align}
    1 = \int\mathcal{D} \theta_{q-1}'\,G_q[B[\theta_{q-1}]]\,\exp\!\left(-\frac{1}{2}\langle \delta \mathrm{d} \theta_{q-1}',\delta\mathrm{d}\theta_{q-1}'\rangle\right)~.
\end{align}
We can repeat the same steps for $\tilde{B}$ and get a similar expression as above. Taking the difference between the two results gives
\begin{align}
    0 = \int\mathcal{D}\theta_{q-1}
'\,\left(G_q[B[\theta_{q-1}]] - G_q[\tilde{B}[\theta_{q-1}]]\right)\,\exp\!\left(-\frac{1}{2}\langle \delta \mathrm{d} \theta_{q-1}',\delta\mathrm{d}\theta_{q-1}'\rangle\right)~.
\end{align}
This implies that the integrand must vanish, so $G_q[B[\theta_{q-1}]] = G_q[\tilde{B}[\theta_{q-1}]]$ and the gauge-fixing functional $G_q$ is independent of $B$ (and hence $\theta_{q-1}$). We can thus factor it out of the integral and take its inverse to find
\begin{align}
    G_q^{-1} &= \int\mathcal{D}\theta_{q-1}\,\exp\left(-\frac{1}{2}\langle \delta B[\theta_{q-1}],\delta B[\theta_{q-1}]\rangle\right) \\
    &=\int\mathcal{D}\theta_{q-1}\,\exp\left[-\frac{1}{2}\left(\langle \delta B,\delta B\rangle-2\langle \delta B,\Box  \theta_{q-1}\rangle-\langle \mathrm{d} \theta_{q-1},\Box \mathrm{d} \theta_{q-1}\rangle\right)\right] \\
    &= \int\mathcal{D}\theta_{q-1}\,\exp\left(+\frac{1}{2}\langle\mathrm{d}\theta_{q-1},\Box\mathrm{d}\theta_{q-1}\rangle\right)~, \label{eq:reduced_G_q}
\end{align}
where in the last equality, we have set $B=0$ for convenience, knowing that both sides are independent of $B$ anyway. If $q>1$ then $G_q^{-1}$ itself has a gauge redundancy $\theta_{q-1} \to \theta_{q-1}[\theta_{q-2}] = \theta_{q-1} + \mathrm{d}\theta_{q-2}$ which we must also account for. Thus, we need to fix the gauge of our gauge-fixing function. 

Seeing that Eq.~\eqref{eq:reduced_G_q} resembles the original path integral with an additional factor of $-\Box$, we define the gauge-for-gauge-fixing functional
\begin{align}
    G^{-1}_{q-1} &= \int\mathcal{D} \theta_{q-2}\, \exp\!\left(+\frac{1}{2}\langle \delta\theta_{q-1}[\theta_{q-2}],\Box\delta\theta_{q-1}[\theta_{q-2}]\rangle\right) \\
    &= \int\mathcal{D} \theta_{q-2}\, \exp\!\left(+\frac{1}{2}\Big[\langle\delta\theta_{q-1},\Box\delta\theta_{q-1}\rangle -2 \langle\delta\Box\theta_{q-1},\Box\theta_{q-2}\rangle - \langle\Box\mathrm{d}\theta_{q-2},\Box\mathrm{d}\theta_{q-2}\rangle\Big]\right)~,
\end{align}
which is similarly independent of $\theta_{q-1}$. We insert $G_{q-1}^{-1}G_{q-1}=1$ into Eq.~\eqref{eq:reduced_G_q} and, just like in the path integral, we factor out an overall gauge-equivalent volume. More explicitly, we start with
\begin{align}
    G_{q}^{-1} &= \int\mathcal{D}\theta_{q-1}\mathcal{D}\theta_{q-2}\,G_{q-1}\exp\!\bigg(+\frac{1}{2}\Big[\langle \mathrm{d}\theta_{q-1},\Box \mathrm{d}\theta_{q-1}\rangle +\langle\delta\theta_{q-1},\Box\delta\theta_{q-1}\rangle \\
    &\hspace{5.5cm} -2 \langle\delta\Box\theta_{q-1},\Box\theta_{q-2}\rangle -\langle\Box\mathrm{d}\theta_{q-2},\Box\mathrm{d}\theta_{q-2}\rangle\Big]\bigg) ~.\nonumber
\end{align}
This can be rewritten as
\begin{align}
    G_{q}^{-1}= G_{q-1}\int\mathcal{D}\theta_{q-1}\mathcal{D}\theta_{q-2}\,\exp\!\bigg(+\frac{1}{2}\Big[ 
     \langle \mathrm{d}\theta_{q-1}\big[\theta_{q-2}\big], \Box\mathrm{d}\theta_{q-1}\big[\theta_{q-2}\big]\rangle + \langle\delta\theta_{q-1}\big[\theta_{q-2}\big],\Box\delta\theta_{q-1}\big[\theta_{q-2}\big]\rangle\Big]\bigg)~.
\end{align}
Next we perform a change of variables $\theta_{q-1}\rightarrow \theta_{q-1}'=\theta_{q-1}[\theta_{q-2}]$ to get
\begin{align}
    G_{q}^{-1}&= G_{q-1}\int\mathcal{D}\theta'_{q-1}\mathcal{D}\theta_{q-2}\,\exp\!\bigg(-\frac{1}{2}  \langle \theta'_{q-1}, \Box^2\theta'_{q-1}\rangle\bigg)\\
    &= \left(\int\mathcal{D}\theta_{q-2}\right) \left(\det\Box_{q-1}\right)^{-1}G_{q-1}~.
\end{align}
Inverting the above relation gives us an expression for $G_{q}$. If $q>2$, then again $G_{q-1}^{-1}$ has its own gauge redundancy, and we must proceed recursively until we reach the scalar case $\theta_{0}$, which has no gauge invariance. For a general layer, we define
\begin{align}
    G_n^{-1} &\equiv \int\mathcal{D}\theta_{n-1}\,\exp\left(-\frac{1}{2}\langle\delta\theta_{n}[\theta_{n-1}],(-\Box)^{q-n}\delta\theta_{n}[\theta_{n-1}]\rangle\right)~,
\end{align}
for $n=1,\cdots ,q$ with $\theta_q\equiv B$ and the terminating condition $G_0 = 1$. As before, we can expand this as
\begin{align}
    G_n^{-1} &= \int\mathcal{D}\theta_{n-1}\,\exp\!\bigg((-1)^{q-n+1}\frac{1}{2}\Big[\langle\delta\theta_{n},\Box^{q-n}\delta\theta_{n}\rangle-2\langle\delta\theta_{n},\Box^{q-n+1}\theta_{n-1}\rangle-\langle\mathrm{d}\theta_{n-1},\Box^{q-n+1}\mathrm{d}\theta_{n-1}\rangle\Big]\bigg)~,\label{eq:G_n_expanded} \\
    &= \int\mathcal{D}\theta_{n-1}\,\exp\!\bigg((-1)^{q-n}\frac{1}{2}\Big[\langle\mathrm{d}\theta_{n-1},\Box^{q-n+1}\mathrm{d}\theta_{n-1}\rangle\Big]\bigg)~,\label{eq:G_n_contracted}
\end{align}
where in the last line we use the fact that $G_n$ is independent of $\theta_n$.
Then using Eq.~\eqref{eq:G_n_contracted} and Eq.~\eqref{eq:G_n_expanded} for $G_n^{-1}$ and $G_{n-1}G_{n-1}^{-1}=1$ respectively, we find
\begin{align}
    G_n^{-1}&= G_{n-1}\int\mathcal{D}\theta_{n-1}\mathcal{D}\theta_{n-2}\,\exp\!\bigg((-1)^{q-n}\frac{1}{2}\Big[\langle\mathrm{d}\theta_{n-1},\Box^{q-n+1}\mathrm{d}\theta_{n-1}\rangle+\langle\delta\theta_{n-1},\Box^{q-n+1}\delta\theta_{n-1}\rangle\\
    &\hspace{6.5cm}-2\langle\delta\theta_{n-1},\Box^{q-n+2}\theta_{n-2}\rangle-\langle\mathrm{d}\theta_{n-2},\Box^{q-n+2}\mathrm{d}\theta_{n-2}\rangle\Big]\bigg)~,\nonumber
\end{align}
which can be rewritten as
\begin{align}
    G_n^{-1}&= \left(\int\mathcal{D}\theta_{n-2}\right)G_{n-1}\int\mathcal{D}\theta_{n-1}\,\exp\!\bigg((-1)^{q-n}\frac{1}{2}\langle\theta_{n-1},\Box^{q-n+2}\theta_{n-1}\rangle\bigg)\\
    &= \left(\int\mathcal{D}\theta_{n-2}\right)G_{n-1}(\det\Box_{n-1})^{(q-n+2)/2}~.
\end{align}
Unpacking this recursive relation, we find that
\begin{align}
    G_q = V_{\mathrm{gauge}}\prod_{k=1}^{q}\left(\det\Box_{q-k}\right)^{(-1)^{k+1}(k+1)/2}~,
\end{align}
where $V_{\mathrm{gauge}}$ is an infinite gauge volume factor which we absorbed into the normalisation of the path integral. This is exactly the tower of ghosts we claimed in Eq.~\eqref{eq:G_q_tower_ghosts}.

\section{Hodge theory and $L^2$ cohomology}\label{app:hodge theory}

Let $M$ be a manifold. Recall that a $k$-form $\alpha$ on $M$ is \textit{closed} if $\mathrm{d}\alpha=0$ and \textit{exact} if $\alpha=\mathrm{d}\beta$ for some $(k-1)$-form $\beta$. Because $\mathrm{d}^2=0$, every exact form is closed, but not every closed form is exact. The failure of this exactness is measured by the \textit{cohomology group} of degree $k$, quotient group
\begin{align}
    H^k(M)\equiv\frac{\{\text{closed $k$-forms}\}}{\{\text{exact $k$-forms}\}}~.
\end{align}
This is a homotopy invariant of $M$.

The Poincar\'e lemma states that $H^k(\mathbb{R}^d)\cong 0$, so any closed form is always exact on a topologically trivial local neighbourhood. In general, given a collection of topologically trivial local neighbourhoods $\{U^i\}$ covering $M=\bigcup_{i}U^i$, a closed form $\alpha$ is exact on each neighbourhood, $\alpha\big|_{U^i}=\mathrm{d}\beta^i\big|_{U^i}$, with each $\beta^i$ differing by a closed form where they overlap.

When computing the determinant of the Hodge Laplacian $\Box_k$ acting on $k$-forms, there is a divergent contribution from each zero mode appearing in the path integral. We call such $k$-forms \textit{harmonic}, and denote the space of harmonic $k$-forms by $\mathcal{H}^k(M)$. To regulate this divergence, we must introduce counterterms proportional to the number of such zero modes.

On compact manifolds, the Hodge theorem \cite{bott_differential_1982} tells us that the number of such zero modes is counted by the Betti numbers, which equal the dimensions of the cohomology groups
\begin{align}
    B_k= \dim H^{k}(M)~.
\end{align}
See, for example, Christensen and Duff \cite{christensen_new_1979}. By the Poincar\'e duality, there is a symmetry between harmonic forms in dual degree, $B_k=B_{d-k}$. 

On the other hand, on non-compact manifolds the path integral only receives contributions from zero modes $\omega$ which are $L^2$ normalisable, i.e.~have finite $\langle \omega,\omega\rangle<\infty$. The theory of $L^2$ harmonic forms is in general much more complicated---for example, the $L^2$ cohomology is not homotopy invariant but instead depends on the quasi-isometry class of the metric \cite{dai_introduction_nodate}, and there is generally no symmetry between the $L^2$ Betti numbers in dual degrees. Nonetheless, it has been shown that on geodesically complete manifolds there is an isomorphism between the `reduced' $L^2$ cohomology $L^2\overline{H}{}^k(M)$ and the $L^2$ harmonic forms ${L^2}\mathcal{H}^k(M)$. This has been used to compute the $L^2$ harmonic forms on (for example) the Eguchi--Hansen space \cite{hitchin_l2-cohomology_2000} and the Euclidean Schwarzschild manifold \cite{etesi_geometric_2001}. In this case the $L^2$ Betti numbers, which we defined to be the dimensions of $L^2 \mathcal{H}^k(M)$, remain duality symmetric. It is these modes which enter into the computation of the $k$-form determinants.

We also rely on complete manifolds to ensure that potential boundary terms vanish. In particular, over $L^2$ forms on complete manifolds there is no boundary term disrupting integration by parts, $\langle \mathrm{d}\omega,\chi\rangle=\langle \omega,\delta \chi\rangle$ \cite{gaffney_special_1954}, and $\Box$ is essentially self-adjoint \cite{chernoff_essential_1973}. 

\printbibliography

\end{document}